\documentclass[
  aps,
  prl,
  reprint,
  superscriptaddress
]{revtex4-2}

\usepackage{comment}
\usepackage{amsmath}
\usepackage{amsfonts}
\usepackage{amssymb}
\usepackage{graphicx}
\usepackage{xcolor}
\usepackage{bm}
\usepackage{comment}

\begin{document}

\title{Comment on ``Andreev Reflection to Probe Momentum-Dependent
Spin Polarization in Altermagnet CrSb''}

\author{Igor I. Mazin}

\affiliation{Department of Physics and Astronomy and
Quantum Science and Engineering Center, George Mason University, Fairfax, VA, USA
}

\author{Maxim Khodas}
\affiliation{Racah Institute of Physics, Hebrew University of Jerusalem, Jerusalem 91904, Israel}

\author{Boris Nadgorny}

\affiliation{
Department of Physics and Astronomy,
Wayne State University,
Detroit, Michigan 48201, USA
}

\maketitle
In a recent Letter, Zhang \textit{et al.}~\cite{Zhang2026} claimed to have discovered momentum-dependent spin polarization in CrSb, a material previously proven to be an altermagnet (AM) with a finite spin-splitting~\cite{Yang2025}, 
by point-contact Andreev reflection (PCAR) ~\cite{Soulen1998, Upadhyay1998}, along three crystallographic orientations, $x$ (\={1}\={1}20) and $y$ ({1}0\={1}0) and $z$ (0001).

We show that the claimed Andreev spin polarization (ASP) is fully forbidden by symmetry along two of the three measured directions in the ballistic regime, and in all three directions in the diffusive regime, and is not established by the data presented.

The authors of Ref. \cite{Zhang2026} -- without establishing the ballistic limit -- claimed to follow Ref.~\cite{MazinSP} (their Ref.~43) in calculating the ASP. However, they used an incorrect definition of ballistic ASP (BASP). 
Their expression for $P(\mathbf{n})$ for the interface normal $\mathbf{n}$ takes the modulus of
$N_{\uparrow}(\mathbf{k})-N_{\downarrow}(\mathbf{k})$ at fixed momentum $\mathbf{k}$, on the grounds that the
Andreev selection rule is ``local in the reciprocal space.'' In fact, it is
local in $\mathbf{k}_{\parallel} = \mathbf{k} - \mathbf{n}(\mathbf{n}\cdot \mathbf{k})$ only; the $k_{\perp} = \mathbf{n}\cdot \mathbf{k}$ is not conserved. 
Therefore, their approach severely overestimates ASP and renders large, nearly identical numbers for three different orientations (see Table \ref{tab:asp}) — despite ASP being forbidden by symmetry for two of them.

\begin{figure}[h!]
    \includegraphics[width=.32\linewidth]{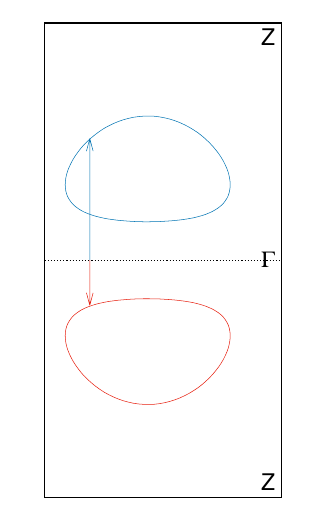}
    \includegraphics[width=.32\linewidth]{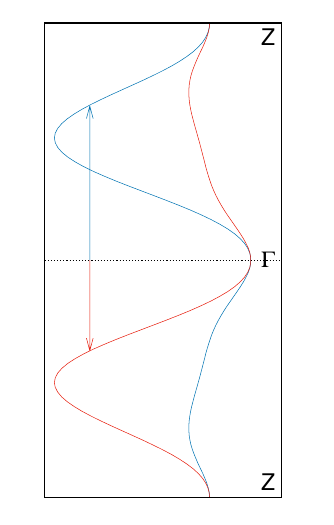}
    \includegraphics[width=.32\linewidth]{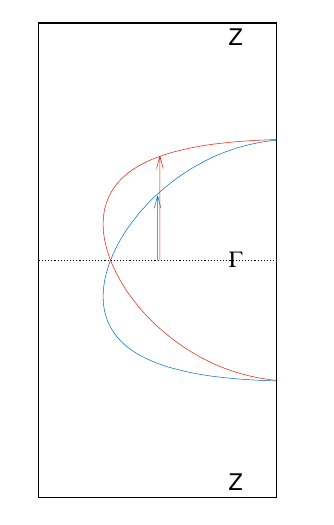}
    \caption{Schematics of three topologically different examples of vertical Fermi surface slices in a g-wave altermagnet. $k_{1\uparrow}>0$ are shown in blue and $k_{2\downarrow}$ in red (note that $v_{Fz}>0$ in all cases, even when $k_2<0$). See the main text for the meaning of $k_{1,2}$.}
    \label{fig}
\end{figure}

Moreover, Ref.~\cite{MazinSP} defines the ``ballistic regime'' by
conservation of $\mathbf{k}_{||}$ across the contact, not merely as a
Sharvin contact \cite{Sharvin}. Partial randomization of
$\mathbf{k}_{||}$ is acceptable in a conventional ferromagnet but
fatal in a $g$-wave AM \cite{Smejkal2022} , whose splitting alternates sign between nodal
sectors. 
Regardless of the regime, the ASP vanishes for
$\mathbf{n}\parallel z$. 
In short, at fixed $\mathbf{k}_{\parallel}$, let $k_1$ and $k_2$ denote the $k_{\perp} =k_z$ of the $\uparrow$ and $\downarrow$
states with outgoing velocities, $v_{\uparrow \downarrow\perp}>0$. Since the dispersion $E_{\sigma}(\mathbf{k})$, $\sigma = \uparrow,\downarrow$, is periodic in
$k_{z}$, the numbers of $v_{\sigma\perp}>0$ and $v_{\sigma\perp}<0$ crossings of the Fermi energy $E_F$
coincide on any sheet, while $[C_{2\perp}\|M_{\mathbf{n}}]$ maps $\uparrow$ at $k_z$ onto $\downarrow$ at $-k_z$; hence, the two spin channels supply equal numbers of outgoing states, whatever the sheet topology
(Fig.~\ref{fig}, where such pairs $k_1,k_2$ are shown). 

In more detail, 
the ASP vanishes for $\mathbf{n}$ such that $[C_{2\perp}||M_{\mathbf{n}}]$ is a symmetry. 
Let $N^{s}_{\sigma}(\mathbf{k}_\parallel)$ count the number of distinct $k_\perp$ satisfying $E_{\sigma}(\mathbf{k}_\parallel,k_\perp)=E_F$ at fixed $\mathbf{k}_\parallel$ under the condition $\mathrm{sign}(\mathbf{v}_\sigma \cdot \mathbf{n}) = s$, $s=+,-$.
For each $\mathbf{k}_\parallel$, Andreev reflection is kinematically
unobstructed if $N^{s}_{\sigma}(\mathbf{k}_\parallel) = N^{-s}_{-\sigma}(-\mathbf{k}_\parallel)$. 
The ASP vanishes if this holds for all $\mathbf{k}_\parallel$.
Taking $\sigma=\uparrow$, $s=+$, by $[C_{2\perp}||M_{\mathbf{n}}]$
symmetry $N^{+}_\uparrow (\mathbf{k}_\parallel)=N^-_\downarrow
(\mathbf{k}_\parallel)$. Periodicity in $k_\perp$ makes upward and downward
crossings of $E_F$ equal in number, so $N^-_\downarrow
(\mathbf{k}_\parallel) =N^+_\downarrow (\mathbf{k}_\parallel)$. 
Finally, in any nonrelativistic collinear magnet, including any AM, $E_\downarrow(\mathbf{k}) = E_\downarrow(-\mathbf{k})$, hence $N^+_\downarrow (\mathbf{k}_\parallel) = N^-_\downarrow (-\mathbf{k}_\parallel)$, and at the end  $N^{+}_{\uparrow}(\mathbf{k}_\parallel) = N^{-}_{\downarrow}(-\mathbf{k}_\parallel)$ for this direction.

In summary, three ingredients are needed for the BASP to vanish for a given direction: a spin-flipping mirror perpendicular to the current direction $\mathbf n$, periodicity in $k_\perp$ and the usual symmetry $E_\sigma (\mathbf{k}) = E_\sigma (-\mathbf{k})$.\footnote{This is consistent with the recent result regarding Andreev reflection in $d$-wave AMs\cite{Sun2023}}
In CrSb, thus, 
BASP is forbidden for $\mathbf{n}\parallel y$ or $z$. 
For $x$ it is allowed. 
Because the correct formula for BASP is \emph{non-local}, it is rather nontrivial to calculate BASP in this case. Our best estimates from the calculated band structure are $\sim 50$\%.

In the diffusive regime, as was pointed out in the original theoretical paper \cite{MazinAR}, AR probes the difference between the effective carrier concentrations in the two spin channels, $\langle N_\sigma(0)(\mathbf{v}_\sigma\cdot \mathbf{n})^2\rangle$, where $\mathbf{n}$ is the transport direction.
At a given spin orientation, $\langle N_\sigma(0)(\mathbf{v}_\sigma\cdot \mathbf{n})^2\rangle$ is a bilinear function of the components of the vector $\mathbf{n}$ that is invariant under the site symmetry subgroup $D_{3d}$ in CrSb.
Therefore, $\langle N_\sigma(0)(\mathbf{v}_\sigma\cdot \mathbf{n})^2\rangle = a_{\sigma} n_z^2 + b_{\sigma} (n_x^2 + n_y^2)$.
Here $a_\sigma$ and $b_\sigma$ are related by the operations in $D_{6h}$ that are not in $D_{3d}$.
Taking, e.g., the screw axis $[C_{2\perp} || C_6]$ immediately gives $a_{\sigma} = a_{-\sigma}$ and $b_{\sigma} = b_{-\sigma}$. Thus, diffusive ASP in CrSb is \textit{identically zero} for all directions.  
Our conclusions are summarized in Table \ref{tab:asp}.
\begin{table}[b]
\caption{\label{tab:asp}%
ASP in CrSb along the three high-symmetry directions measured in
Ref.~\cite{Zhang2026}, compared with the values reported there.}
\begin{ruledtabular}
\begin{tabular}{lcccc}
 & \multicolumn{2}{c}{This work} & \multicolumn{2}{c}{Ref.~\cite{Zhang2026}} \\
\cline{2-3}\cline{4-5}
$\mathbf{n}$ & Ballistic & Diffusive & calc. & exp. \\
\colrule
$z$ $(0001)$             & 0                            & 0 & 87.4\% & 73.4\% \\
$x$ $(\bar{1}\bar{1}20)$ & $\sim\!50\%$\footnotemark[1] & 0 & 85.6\% & 67.9\% \\
$y$ $(10\bar{1}0)$       & 0                            & 0 & 85.4\% & 61.9\% \\
\end{tabular}
\end{ruledtabular}
\footnotetext[1]{Band-structure estimate in the $Z\to0$ ballistic limit; all other
entries in the first two columns are exact.} 
\end{table}

Furthermore, the data analysis is flawed. As we noted, ballistic transport in Ref.~\cite{Zhang2026} is not demonstrated; contact resistances span almost two orders of magnitude, yet neither $\rho$, $\ell$, nor the Sharvin contact size is given ~\cite{Sharvin,Wexler,Woods2004,Zutic,Chalsani2007, NadgornyCh}. 
The BTK expression applies to normalized conductance, but the normalization of the measured spectra--which materially affects the extracted ASP \cite{Bugoslavsky2005, Woods2004} is not described. The minimized ``area difference'' $D$ is given no definition, weighting, or uncertainty, and its rescaling
$D_{N}=(D_{i}-D_{\min})/(D_{\max}-D_{\min})$ maps the extrema to zero
and one irrespective of absolute fit quality. The unpolarized fits in Fig.~2 require $Z=5.7$, $0.95$, and $0.90$; only the first is claimed to be unphysical. No reduced $\chi^{2}$, residual analysis, or uncertainties on $P$ are reported: neither the improvement over $P=0$ nor the differences among the quoted $73.4\%$, $67.9\%$, and $61.9\%$ can be assessed.
The formalism \cite{CTC} used in Ref.~\cite{Zhang2026} does not conserve
charge~\cite{Eschrig2013}; while
this alone cannot account for the reported values, it is an
additional source of error.

We stress that any mechanism that can make the measurement non-zero (misorientation, strain, spin-dependent scattering amplitudes) is a property of the contact itself and severs the correspondence between the observation and the putative spin polarization at a fixed interface orientation.

\bibliography{bib}

\end{document}